\documentclass{PoS}

\title{Dissipative hydrodynamic evolution of hot quark matter at finite baryon density}

\ShortTitle{Dissipative hydrodynamic evolution of hot quark matter at finite baryon density}

\author{\speaker{Akihiko Monnai}\\
       Department of Physics, The University of Tokyo, Tokyo 113-0033, Japan\\
       Theoretical Research Division, Nishina Center, RIKEN, Wako 351-0198, Japan\\
       E-mail: \email{monnai@nt.phys.s.u-tokyo.ac.jp}}

\abstract{High-energy heavy ion collider experiments at RHIC and LHC have revealed that relativistic hydrodynamic models describe the hot and dense quark matter quantitatively. In this study, I develop a novel dissipative hydrodynamic model at finite baryon density to investigate the net baryon rapidity distribution. The results show that the distribution is widened in hydrodynamic evolution, which implies that the transparency of the collisions is effectively enhanced. This suggests that the kinetic energy loss for medium production at the initial stage could be larger. Furthermore, the net baryon distribution is found sensitive to baryon diffusion, implying that dissipative hydrodynamic modeling would be important for understanding the hot medium.
}

\FullConference{Xth Quark Confinement and the Hadron Spectrum,\\
		October 8-12, 2012\\
		TUM Campus Garching, Munich, Germany}

\begin{document}

\section{Introduction}

Quarks and gluons are considered to be deconfined from hadrons at high temperatures and to form the quark-gluon plasma (QGP) \cite{Yagi:2005yb}. It had been na\"{i}vely expected that the plasma is weakly coupled because of the asymptotic freedom of QCD. High-energy heavy ion collisions revealed, on the contrary, that the hot QCD matter would be strongly coupled near the quark-hadron crossover. The experimentally observed hadronic yields turned out to be described quantitatively by relativistic hydrodynamic models with small viscosity at the Relativistic Heavy Ion Collider (RHIC) and the Large Hadron Collider (LHC). Modern hydrodynamic analyses take account of viscosity and fluctuation to improve the model accuracy \cite{Schenke:2010rr}. However, the net baryon number is neglected in most of those studies even though it would be important in understanding the experimental results such as particle-to-antiparticle ratios. This is possibly because viscous hydrodynamic equations of motion are complicated in the presence of finite charge density and also the boost invariant ansatz is often assumed, by which the net baryon number conserved at forward rapidity is omitted. 

Baryon stopping has been used to quantify the transparency of collisions \cite{Busza:1983rj, Videbaek:1995mf}. At early fixed-target experiments in the Alternating Gradient Synchrotron (AGS) and the Super Proton Synchrotron (SPS), the mean rapidity of the baryon number is largely reduced from that of the incoming projectile, suggesting that the nuclei stay near the center of mass after the collision. The later collider experiments at RHIC and LHC, on the other hand, show that the collisions become increasingly transparent at high energies. This na\"{i}vely implies that only part of energy is used for the production of a hot and dense medium for the latter cases. In this study, I investigate the effects of dissipative hydrodynamic medium interaction at the later stage on the net baryon rapidity distribution to find that more energy could be available for the hot medium \cite{Monnai:2012jc,Monnai:2013rz}.

Development of a finite-density dissipative hydrodynamic model would also be useful in the search of the critical point on the phase diagram of QCD because the first principle approaches have difficulty at non-vanishing chemical potential due to the sign problem. The applicability of the hydrodynamic picture at lower energies would need investigating through comparison of theoretical calculations and experimental data since the beam energy scan at RHIC show that large momentum azimuthal anisotropy would stay relatively large \cite{Aggarwal:2010cw}. In such systems dissipative process, which accounts for the deviation from local equilibrium, would become important. The beam energy scans are also planned and proposed at Facility for Antiprotons and Ion Research, Nuclotron-based Ion Collider fAcility and Japan Proton Accelerator Research Complex.

\section{Finite-density dissipative hydrodynamics in relativistic systems}

Dissipative processes in relativistic systems are non-trivial since the relativistic extension of Navier-Stokes equation is subject to acausality and instablity. The equation is obtained by introducing off-equilibrium correction to the entropy current up to the first order in terms of dissipative currents. On the other hand, one can consider the next order corrections that introduce relaxation effects to the theory. This second order theory is known to be causal and stable and is widely used for the analyses of dissipative phenomena in relativistic fluids. In this study I employ the Israel-Stewart formulation \cite{Israel:1979wp} extended for the systems with particle number changing processes \cite{Monnai:2010qp} as the foundation of the causal dissipative hydrodynamic model.

The flow tensor decomposition of the conserved quantities yields thermodynamic variables; the energy density $e_0 = u_\mu u_\nu T_0^{\mu \nu} $, the pressure $P_0 = -\Delta_{\mu \nu} T_0^{\mu \nu}/3$ and the net baryon number $n_{B0} = u_\mu N_{B0}^\mu$. The dissipative currents are the bulk pressure $\Pi = -\Delta_{\mu \nu} \delta T^{\mu \nu}/3$, the shear stress tensor $\pi^{\mu \nu} = [(\Delta^{\mu}_{\ \alpha} \Delta^{\nu}_{\ \beta} + \Delta^{\mu}_{\ \beta} \Delta^{\nu}_{\ \alpha})/2 - \Delta^{\mu \nu} \Delta_{\alpha \beta}/3]\delta T^{\alpha \beta} = \delta T^{\langle \mu \nu \rangle}$, the baryon dissipation current $V^\mu = \Delta^{\mu}_{\ \nu} V_B^\nu$. Here $u^\mu$ is the flow and $\Delta^{\mu \nu} = g^{\mu \nu} - u^\mu u^\nu$ is the projection operator. Energy-momentum and baryon number conservations read
\begin{eqnarray}
\partial_\mu T^{\mu \nu} &=& u^\nu [D e_0 + (e_0 + P_0 + \Pi ) \nabla _\mu u^\mu - \pi^{\mu \nu} \nabla _{\langle \mu} u_{\nu \rangle} ] \nonumber \\
&+& (e_0 + P_0 + \Pi ) D u^\mu - \nabla ^\mu (P_0 + \Pi )  - \pi^{\mu \nu} D u_\nu + \Delta ^{\mu \nu} \nabla ^\rho \pi_{\nu \rho} = 0, \\
\partial_\mu N_B^\mu &=& D n_{J0} + n_{J0} \nabla _\mu u^\mu + \nabla _\mu V_J^\mu - V_J^\mu D u_\mu = 0.
\end{eqnarray}
Here $D = u^\mu \partial_\mu$ and $\nabla^\mu = \Delta^{\mu \nu} \partial_\nu$ reduce to the time and the space derivatives in the local rest frame. The entropy production can be written in the extended second order theory \cite{Monnai:2010qp} as
\begin{eqnarray}
\partial_\mu s^\mu = - \Pi X_\Pi - V_\mu X^\mu_V + \pi_{\mu \nu} X^{\mu \nu}_\pi \geq 0,
\end{eqnarray}
where the thermodynamic forces with the next-to-leading order corrections are
\begin{eqnarray}
X_\Pi &=& \nabla _\mu u^\mu - \frac{1}{\zeta_{\Pi \Pi}} \bigg[ - \zeta_{\Pi \delta e} D\frac{1}{T} + \zeta_{\Pi \delta n_B} D\frac{\mu_B}{T} - \tau_\Pi D \Pi \nonumber \\
&+& \chi_{\Pi \Pi}^{a} \Pi D\frac{\mu _B}{T} + \chi_{\Pi \Pi}^b \Pi D \frac{1}{T} + \chi_{\Pi \Pi}^c \Pi \nabla _\mu u^\mu + \chi_{\Pi V}^{a} V_\mu \nabla ^\mu \frac{\mu_B}{T} + \chi_{\Pi V}^b V_\mu \nabla ^\mu \frac{1}{T} \nonumber \\
&+& \chi_{\Pi V}^c V_\mu D u ^\mu + \chi_{\Pi V}^d \nabla ^\mu V_\mu + \chi_{\Pi \pi} \pi _{\mu \nu} \nabla ^{\langle \mu} u^{\nu \rangle} \bigg] , 
\\
X_V^\mu &=& \nabla ^\mu \frac{\mu _B}{T} + \frac{1}{\kappa_{V}} \bigg[ - \kappa _{V W} \bigg( \nabla ^\mu \frac{1}{T} + \frac{1}{T} D u^\mu \bigg) - \tau _{V} \Delta^{\mu \nu} D V_\nu \nonumber \\
&+& \chi_{V V}^{a} V^\mu D \frac{\mu _B}{T} + \chi_{V V}^b V^\mu D \frac{1}{T} + \chi_{V V}^c V^\mu \nabla _\nu u^\nu + \chi_{V V}^d V^\nu \nabla _\nu u^\mu + \chi_{V V}^e V^\nu \nabla ^\mu u_\nu \nonumber \\
&+& \chi_{V \pi}^{a} \pi^{\mu \nu} \nabla_\nu \frac{\mu _B}{T} + \chi_{V \pi}^b \pi^{\mu \nu} \nabla_\nu \frac{1}{T} + \chi_{V \pi}^c \pi ^{\mu \nu} D u_\nu + \chi_{V \pi}^d \Delta^{\mu \nu} \nabla ^\rho \pi _{\nu \rho} \nonumber \\
&+& \chi_{V \Pi}^{a} \Pi \nabla ^\mu \frac{\mu _B}{T} + \chi_{V \Pi}^b \Pi \nabla ^\mu \frac{1}{T} + \chi_{V \Pi}^c \Pi D u^\mu + \chi_{V \Pi}^d \nabla ^\mu \Pi \bigg] , \label{V}\\
X_\pi^{\mu \nu} &=& \nabla ^{\langle \mu} u^{\nu \rangle} + \frac{1}{2 \eta} \bigg[ - \tau_\pi D \pi^{\langle \mu \nu \rangle}  \label{eq:XV} \nonumber \\
&+& \chi _{\pi \Pi} \Pi \nabla ^{\langle \mu} u^{\nu \rangle} + \chi_{\pi \pi}^{a} \pi^{\mu \nu} D \frac{\mu_B}{T} + \chi_{\pi \pi}^b \pi^{\mu \nu} D \frac{1}{T} + \chi_{\pi \pi}^c \pi ^{\mu \nu} \nabla _\rho u^\rho + \chi_{\pi \pi}^d \pi ^{\rho \langle \mu} \nabla _\rho u^{\nu \rangle} \nonumber \\
&+& \chi_{\pi V}^{a} V^{\langle \mu} \nabla ^{\nu \rangle} \frac{\mu_B}{T} + \chi_{\pi V}^b V^{\langle \mu} \nabla ^{\nu \rangle} \frac{1}{T} + \chi_{\pi V}^c V^{\langle \mu} D u^{\nu \rangle} + \chi_{\pi V}^d \nabla ^{\langle \mu} V ^{\nu \rangle} \bigg] .
\end{eqnarray}
Here $T$ is the temperature and $\mu_B$ is the baryon chemical potential. The semi-positive definiteness of the entropy production suggests $\Pi = -\zeta_{\Pi \Pi} X_\Pi$, $V^\mu = \kappa_{V} X_V^\mu$ and $\pi^{\mu \nu} = 2\eta X_\pi$ where $\zeta_{\Pi \Pi}$, $\kappa_V$ and $\eta$ are the (bare) bulk viscosity, the baryon charge conductivity and the shear viscosity, respectively. The linear cross coefficients -- the bulk-energy and the bulk-baryon cross viscosities $\zeta_{\Pi \delta e}$ and $\zeta_{\Pi \delta n_B}$ and the baryon-heat conductivity $\kappa_{VW}$ -- can be either positive or negative. The linear response theory require that the coefficients satisfy Onsager reciprocal relations \cite{Onsager, Onsager2}, a representation of microscopic reversibility. $\tau_\Pi$, $\tau_V$ and $\tau_\pi$ are the relaxation times and $\chi$'s are the second order transport coefficients. 

Since the main focus is on time-evolution of the system in the rapidity direction as the net baryon number originates from the shattered nuclei which resides at forward rapidity, transverse dynamics is simply integrated out in the model. This would be justified for the net baryon distribution because the dependence of the net baryon yields on transverse geometry is experimentally shown to be very small \cite{Adler:2003cb}.

The equation of state (EoS) and the transport coefficients are necessary to solve the hydrodynamic equations of motion. Latest results of lattice QCD calculations for the EoS in the vanishing chemical potential \cite{Borsanyi:2010cj} and the quadratic baryon fluctuation \cite{Borsanyi:2011sw} are used for constructing the finite-density EoS in Taylor expansion method. Due to difficulties in first principle calculations of the dynamical responses, the viscosity and the conductivities are chosen as $\eta = s/4\pi$ \cite{Kovtun:2004de}, $\zeta = 5 [(1/3) - c_s^2] \eta$ \cite{Hosoya:1983id, Monnai:2012jc} and $\kappa_V = c_V (\partial \mu_B / \partial n_B)^{-1}_T / 2\pi$ \cite{Natsuume:2007ty}. Here $s$ is the entropy density and $c_S$ is the sound velocity. They are based on Anti-de Sitter/conformal field theory correspondence except for the bulk viscosity, which is non-conformal and is estimated here with the method based on the non-equilibrium statistical operator method. Note that the scalar linear cross terms are merged into this effective bulk viscosity. The baryon-heat cross conductivity is parametrized as $\kappa_{VW} = c_{VW} [n_{B0}/(e_0+P_0)] (5\eta T \kappa_V)^{1/2}$ from the argument based on matter-antimatter symmetry, semi-positive definiteness of the transport coefficient matrix and dimensionality \cite{Monnai:2012jc}. $c_V$ and $c_{VW}$ are parameters. The relaxation times are set to $\tau_\pi = (2 - \ln 2) / 2\pi T$, $\tau_\Pi = 18 - (9\ln 3 - \sqrt{3} \pi) / 24\pi T$ and $\tau_V = \ln 2 / 2\pi T$ according to the string theoretical approach \cite{Natsuume:2007ty}. The other second order transport coefficients are set to vanishing for the moment to see the qualitative responses of the system to each dissipative current more clearly.

The initial conditions are estimated from the color glass theory, a description of gluon saturation. Energy density profile is employed from the gluon distribution in Nara adaptation \cite{Drescher:2006ca,Drescher:2007ax} of Monte-Carlo Kharzeev-Levin-Nardi model \cite{Kharzeev:2002ei,Kharzeev:2004if} by averaging it over the transverse area as $e_0(\tau_0,\eta_s) = (1/ S_\mathrm{area})dE_T/\tau_0 d\eta_s$ where $\eta_s$ is the space-time rapidity. Here the gluon energy distribution is assumed not to be much modified before the initial time $\tau_0$, which is chosen as 1 fm/$c$. Net baryon density profile is obtained from the valence quark distribution in the color glass medium \cite{MehtarTani:2008qg,MehtarTani:2009dv} as $n_{B0}(\tau_0,\eta_s) = (1/S_\mathrm{area})dN_{B-\bar{B}}/\tau_0 d\eta_s$. The dissipative currents are set to vanishing at the initial time to better illustrate the effects of the off-equilibrium processes.

\section{Results}

\begin{figure} 
\includegraphics[width=.48\textwidth]{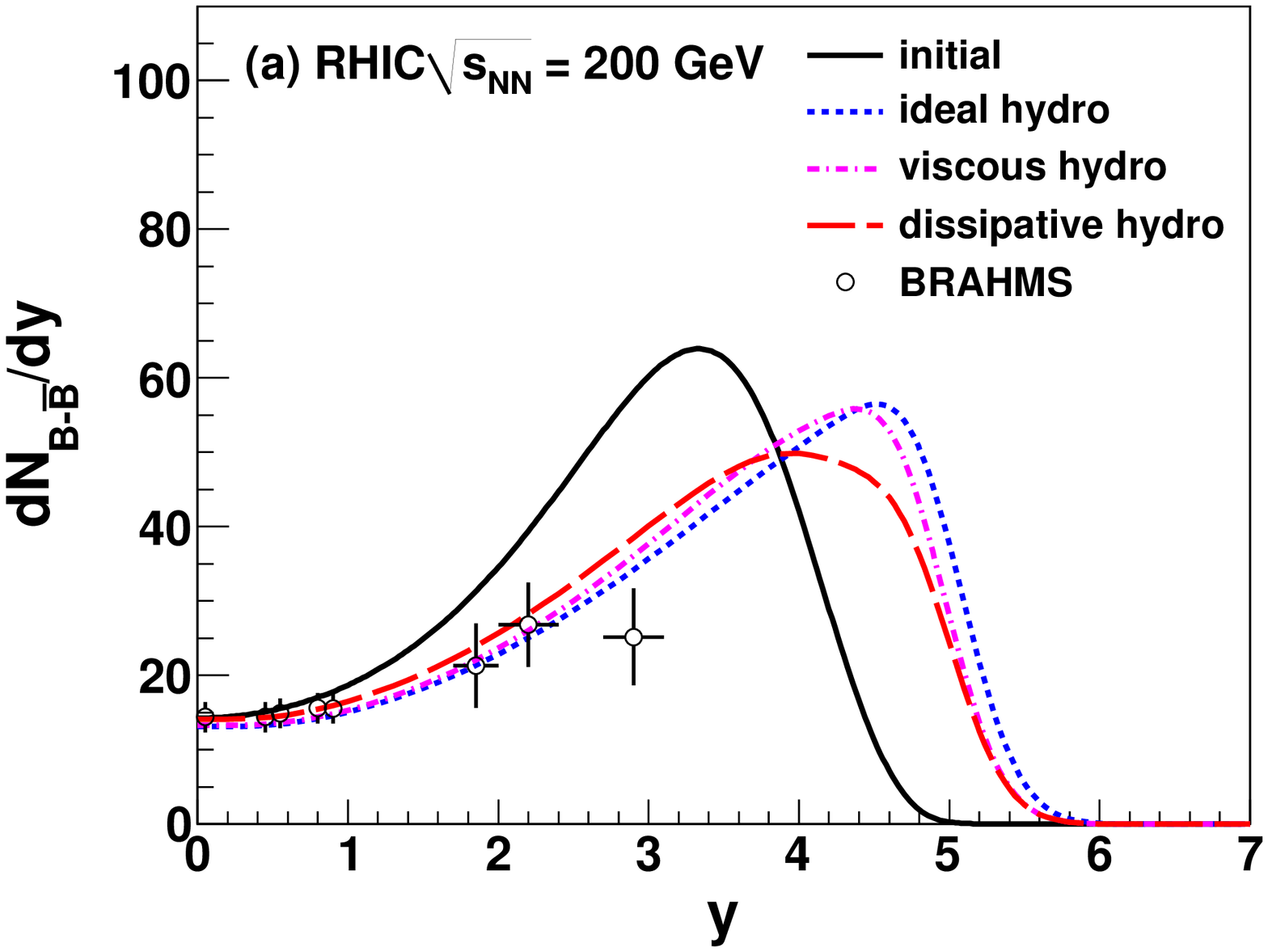} 
\includegraphics[width=.48\textwidth]{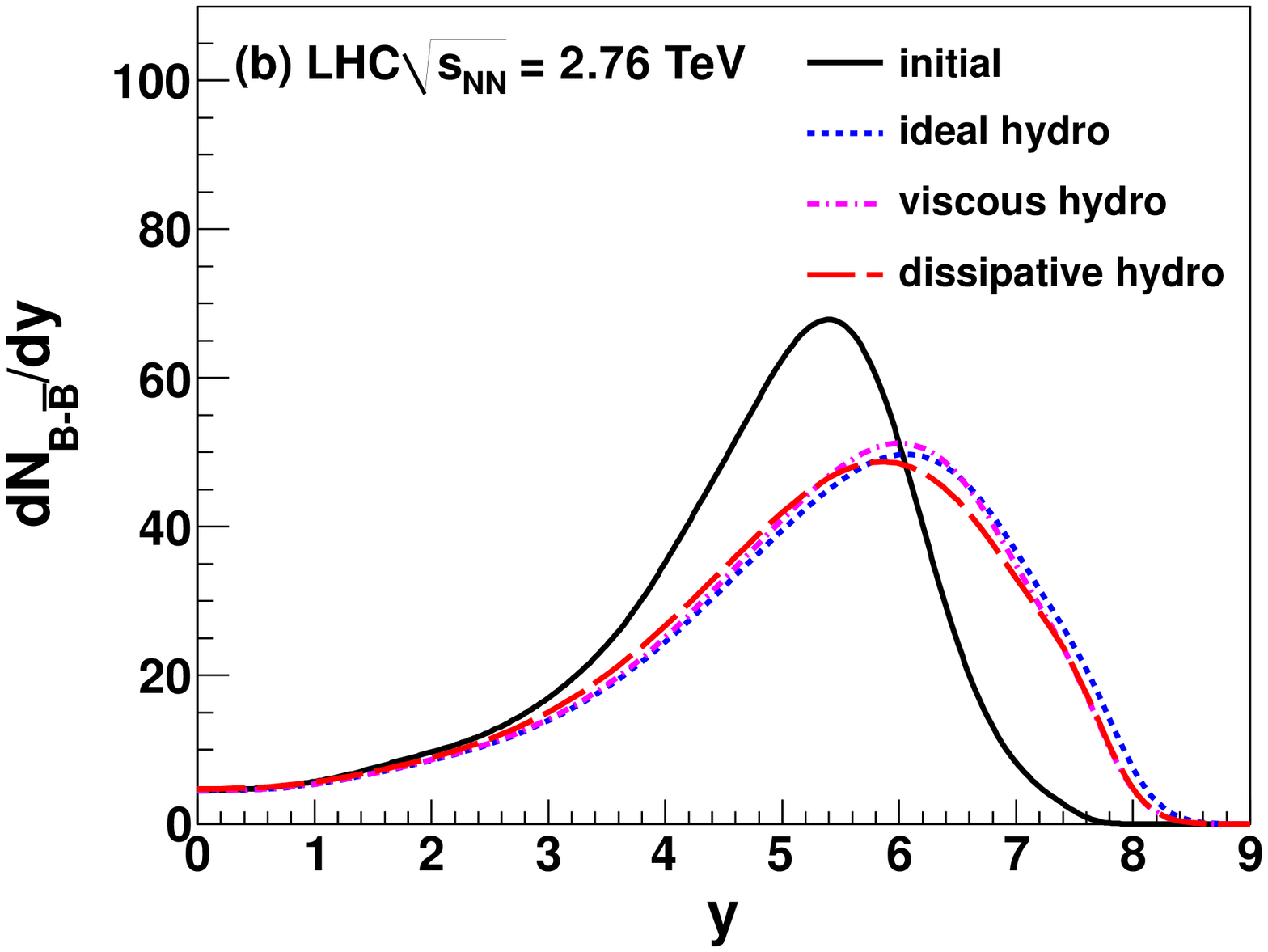} 
\caption{Net baryon rapidity distributions of the color glass-based initial state (solid lines) and of ideal (dashed lines), viscous (dash-dotted lines) and dissipative (dotted lines) hydrodynamic results at (a) RHIC and (b) LHC. The data points are the scaled results of the BRAHMS Collaboration.} 
\label{fig:1} 
\end{figure}

Figure~\ref{fig:1} shows the net baryon rapidity distributions of the color glass-based initial state and the ones after ideal, viscous and dissipative hydrodynamic evolution for Au-Au collisions in RHIC at $\sqrt{s_{NN}} = 200$ GeV and Pb-Pb collisions in LHC at $\sqrt{s_{NN}} = 2.76$ TeV. Here the viscous hydrodynamic results include shear and bulk viscous effects and the dissipative hydrodynamic ones include baryon diffusion effect in addition to the viscous ones. The parameter in the transport coefficients are set to $c_V = 1$ and $c_{VW} = 0$ for the moment. Freeze-out is calculated in Cooper-Frye formula \cite{Cooper:1974mv} with the freeze-out temperature $T_f = 0.16$ GeV. Note that the post-hydrodynamic evolution is not considered here because the modification would be small in dilute media and also the net baryon number does not change in hadronic decay.

One can see that the hydrodynamic evolution widens the net baryon distributions as the outward flow flux is induced by the pressure gradient and that the distribution after the hydrodynamic evolution exhibits overall agreement with the scaled results of the net proton distribution experimentally observed by the BRAHMS Collaboration \cite{Bearden:2003hx} at RHIC. Effects of viscosity and diffusion are suggested to be more visible at RHIC than at LHC. Viscosity tends to lessen the hydrodynamic effect because the longitudinal pressure is reduced as dynamical responses to expansion. Diffusion also steepens the distribution since gradients in the chemical potential field are induced by the net baryon number at forward rapidity. It is note-worthy that this suggest the finite density dissipative processes, which is na\"{i}vely believed to be small, might have visible effects on observables.

\begin{figure} 
\begin{center}
\includegraphics[width=.48\textwidth]{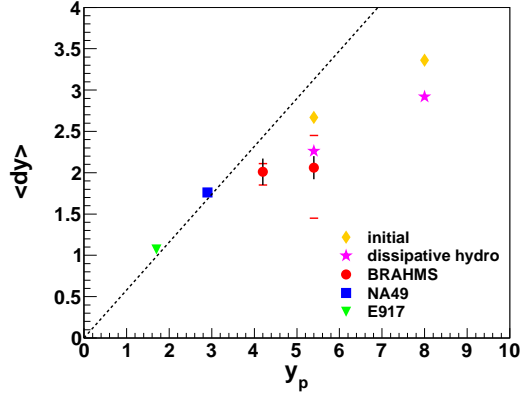} 
\caption{Mean rapidity losses at E917 (AGS) \cite{Back:2000ru}, NA49 (SPS) \cite{Appelshauser:1998yb} and BRAHMS (RHIC) \cite{Bearden:2003hx,Arsene:2009aa} experiments and those of initial distributions and dissipative hydrodynamic results at RHIC and LHC. The dotted line denotes a linear extrapolation of the AGS and SPS results and the bars on the RHIC data show ambiguities due to the lack of experimental data at forward rapidity \cite{Bearden:2003hx}. } 
\label{fig:2} 
\end{center}
\end{figure}

Baryon stopping is quantified with the mean rapidity loss $\langle \delta y \rangle = y_p - \langle y \rangle$ \cite{Busza:1983rj,Videbaek:1995mf}, where $y_p$ is the rapidity of the incoming projectile and 
\begin{equation}
\langle y \rangle = \int _0^{y_p} y \frac{dN_{B-\bar{B}}(y)}{dy} dy  
\bigg/ \int _0^{y_p} \frac{dN_{B-\bar{B}}(y)}{dy} dy.
\end{equation}
The rapidity losses of the initial conditions are $\langle \delta y \rangle_\mathrm = 2.67$ at RHIC and 3.36 at LHC for the current parameter settings. It should be noted they are dependent on parameters in the color glass model and can be chosen differently. Ideal, viscous and dissipative hydrodynamic evolution reduces them to $\langle \delta y \rangle$ = 2.09, 2.16 and 2.26 at RHIC and 2.82, 2.86 and 2.92 at LHC, respectively. The dissipative hydrodynamic results are plotted along with the experimental data from AGS, SPS and RHIC in Fig.~\ref{fig:2}. The simple linear extrapolation of the results from the lower energy collisions deviates from the RHIC data, which implies that collisions become increasingly transparent as the energy becomes higher. The deviation, on the other hand, would be smaller when the hydrodynamic effects are taken into account. This indicates that more energy is available for the production of a hot medium than that implied from the experimental data.

Finally, the effects of the cross terms are investigated. As mentioned earlier, the linear scalar terms are combined into the effective bulk viscosity;
\begin{eqnarray}
\Pi &=& - \zeta_{\Pi \Pi} \frac{1}{T}  \nabla_\mu u^\mu - \zeta_{\Pi \delta_e} D \frac{1}{T} + \zeta_{\Pi \delta n_B} D \frac{\mu_B}{T} + \mathcal{O}(\delta^2) \nonumber \\
&=& - \bigg[ \frac{ \zeta_{\Pi \Pi}}{T} + \frac{\zeta_{\Pi \delta_e}}{T} \bigg( \frac{\partial P_0}{\partial e_0} \bigg)_{n_{B0}} + \frac{\zeta_{\Pi \delta n_B}}{T} \bigg( \frac{\partial P_0}{\partial n_{B0}} \bigg) _{e_0} \bigg] \nabla _\mu u^\mu + \mathcal{O}(\delta^2) \nonumber \\
&\equiv& - \zeta \nabla _\mu u^\mu + \mathcal{O}(\delta^2) .
\label{eq:scalar_relation}
\end{eqnarray}
If one assumes $\zeta_{\Pi \delta {n_B}} = [n_{B0}/(e_0+P_0)] \times \zeta_{\Pi \delta e}$, $\zeta = (\zeta_{\Pi \Pi} + \zeta_{\Pi \delta e} c_s^2)/T$ follows from thermodynamic relations. It should be noted that the response of a system to expansion, \textit{i.e.}, the bare bulk viscosity $\zeta_{\Pi \Pi}$ could be as large as the shear viscosity $\eta$, but cancellation by the cross term would occur except for the crossover region where the sound velocity is small. Phenomenologically, the bare bulk viscous term discourages the expansion but the cross term encourages it in an effort to decrease the temperature/chemical potential of the system, leading to the cancellation of the effects. This could be a reason for the general smallness of the overall bulk viscosity.

\begin{figure} 
\begin{center}
\includegraphics[width=.48\textwidth]{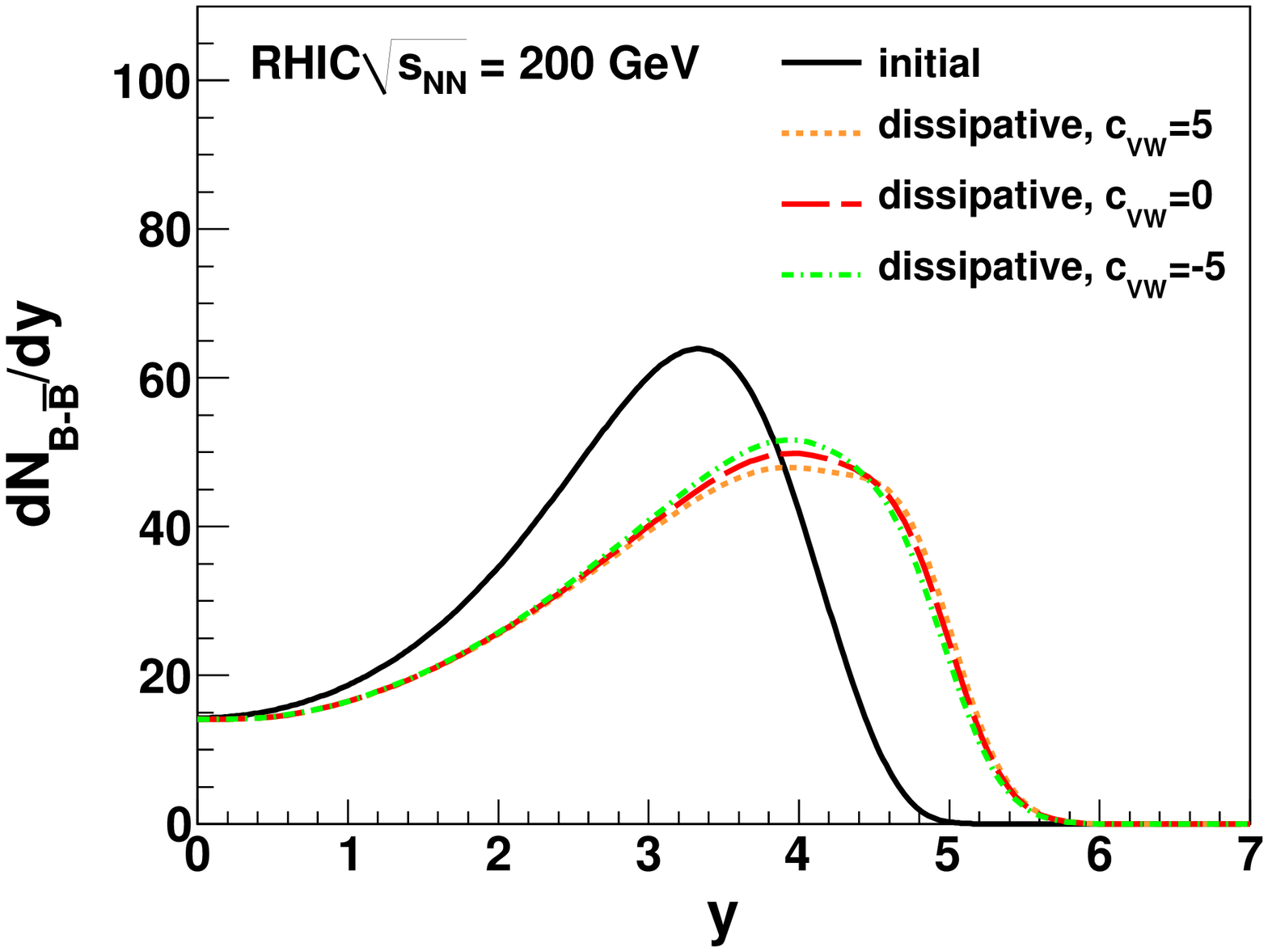} 
\includegraphics[width=.48\textwidth]{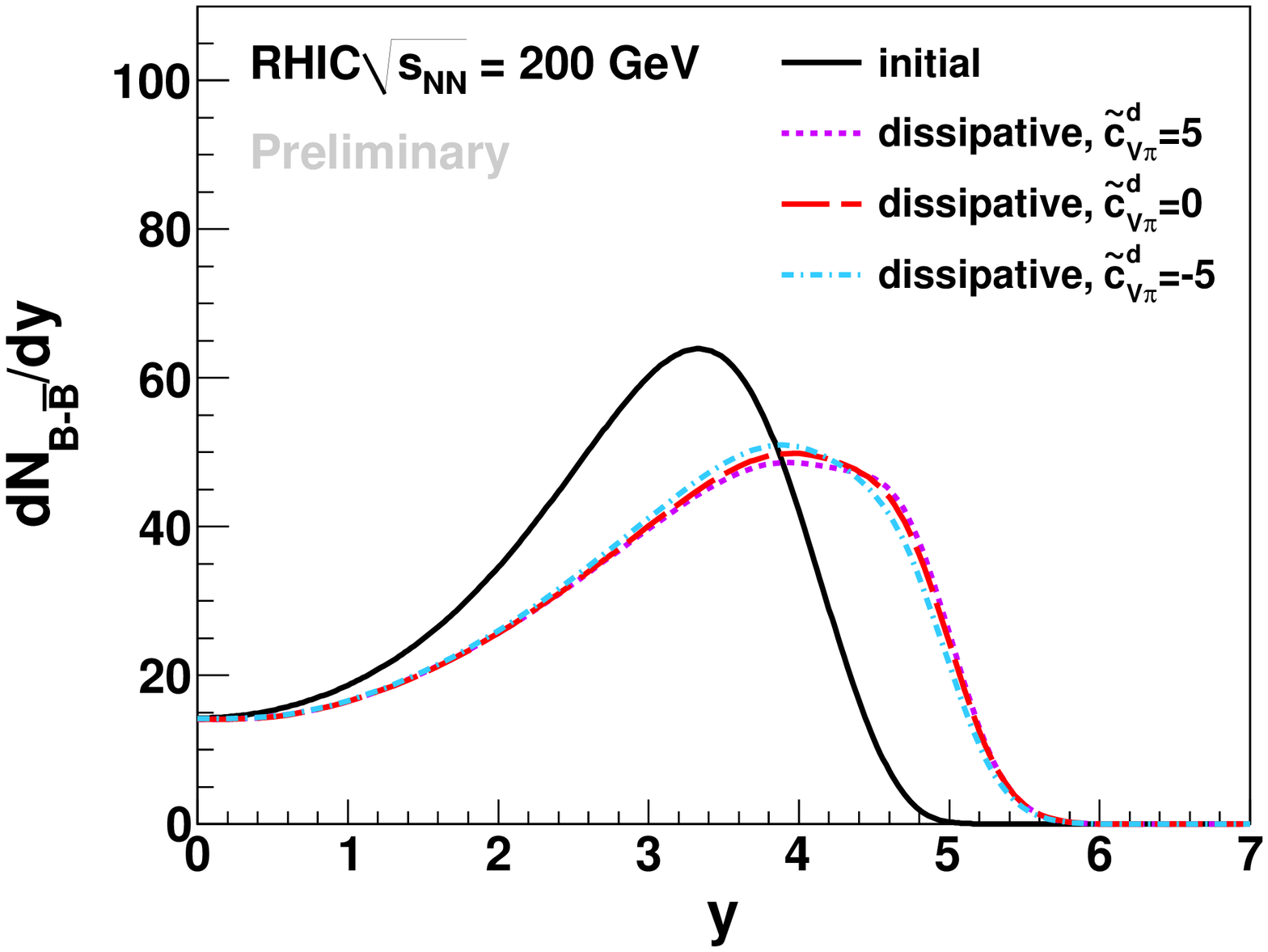} 
\caption{Net baryon rapidity distributions of the color glass-based initial state (solid lines) and of dissipative hydrodynamic results with (a) the linear cross factor $c_{vW} = 5$ (dotted lines), $c_{vW} = 0$ (dashed lines) and $c_{vW} = -5$ (dash-dotted lines) and (b) the second order cross factor $\tilde{c}_{v\pi} = 5$ (dotted lines), $\tilde{c}_{v\pi} = 0$ (dashed lines) and $\tilde{c}_{v\pi} = -5$ (dash-dotted lines) at RHIC.} 
\label{fig:3} 
\end{center}
\end{figure}

The baryon dissipation current, a vector dissipative current, has the thermo-diffusion term as a cross term. Diffusion induced by thermal gradients is known as Soret effect in statistical mechanics. This could be potentially as large as the bare baryon dissipation. The numerical result shows, on the other hand, the effect is rather small for high-energy heavy ion collisions (Fig.~\ref{fig:3}). This is because the cross conductivity should vanish at the vanishing limit of the chemical potential because of the matter-antimatter symmetry condition $V(-\mu_B) = -V(\mu_B)$, limiting Soret effect at forward rapidity where the net baryon number is non-vanishing. The shear viscosity does not have the cross term because it is the only tensor dissipative current. 

The second order cross terms could also be important because the shear stress tensor is generally much larger than the bulk pressure, which in turn is larger than the baryon dissipation current. It is numerically found that the bulk-shear coupling term in $\Pi$ and the shear-baryon and the bulk-baryon coupling terms in $V^\mu$ would potentially be large and should be carefully treated for the second order theory to be valid. For example, Fig. \ref{fig:3} (b) shows the effect of the baryon-shear term $\chi_{V \pi}^d \Delta^{\mu \nu} \nabla ^\rho \pi _{\nu \rho}$ in Eq.~(\ref{eq:XV}) when the transport coefficient is parametrized as $\chi_{V\pi}^d = \tilde{c}_{V\pi}^d [n_{B0}/(e_0+P_0)] \tau_V$ with $\tilde{c}_{V\pi}^d = 5$, 0 and -5. In this case the second order cross term exhibits the trend similar to the linear one. Note that the magnitude of the effects are dependent on the transport coefficients which are in general not well known.

\section{Discussion and conclusions}

In this study, I developed a relativistic dissipative hydrodynamic model at finite baryon density to estimate the net baryon rapidity distribution. The transverse dynamics is integrated out as it is motivated by the experiments \cite{Adler:2003cb}. Numerical estimations are performed with RHIC and LHC settings with the equation of state constructed from lattice QCD calculations in Taylor expansion method. The initial conditions for the energy and the net baryon number distributions are based on the color glass picture. The results suggest that the net baryon number is carried to forward rapidity with the flow during the time-evolution, effectively reducing the mean rapidity loss at both RHIC and LHC. This could explain the rather high transparency of the collisions at RHIC compared with that implied from the simple extrapolation of the AGS and SPS results. They also suggest that more kinetic energy would be converted into the inertial energy of the QGP than previously deduced from the experimental data. The results at RHIC show that the effects of baryon diffusion and viscosity could be visible. The thermo-diffusion effect, or Soret effect, was investigated and found to be limited in the forward rapidity region in high-energy heavy ion collisions due to the matter-antimatter symmetry condition on the cross conductivity. 

Introduction of transverse dynamics and more realistic transport coefficients would increase accuracy of the model. The former would not directly affect the net baryon distribution but it can affect the space-time evolution itself and thus the lifetime of the produced hot medium. The latter would be important because the dissipative effects could be underestimated without the realistic chemical potential dependences of the transport coefficients. It would also be worth-investigating to see the correspondence between the mean rapidity losses for the net baryon number analyzed in this study and the ones for the energy which has recently measured at LHC \cite{Woehrmann}. 

\section*{Acknowledgement}

The author is grateful for the valuable comments by T. Hatsuda. 
The work of A.M. is supported by the Grant-in-Aid for JSPS Research Fellows (No. 10J07647).

\end{document}